\documentclass[12pt]{iopart}
\usepackage{color}
\usepackage{iopams} 
\usepackage{graphicx}
\usepackage{enumerate}

\newcommand{\expdec}{e^{-\tau/\ta}}
\newcommand{\DUt}{\Delta U_\tau}
\newcommand{\Wt}{W_\tau}
\newcommand{\wt}{w_\tau}
\newcommand{\Qt}{Q_\tau}
\newcommand{\qt}{q_\tau}
\newcommand{\Wn}{W_n}
\newcommand{\wn}{w_n}
\newcommand{\Qn}{Q_n}
\newcommand{\qn}{q_n}
\newcommand{\dd}{\mathrm{d}}
\newcommand{\Ieff}{I_{\mathrm{eff}}}
\newcommand{\ta}{\tau_\alpha}

\begin{document}
\title[Fluctuation theorems for harmonic oscillators]{Fluctuation theorems for harmonic oscillators}
\author{S. Joubaud, N. B. Garnier, S. Ciliberto}
\address{Laboratoire de Physique de l'ENS Lyon, CNRS UMR 5672,
       46, All\'ee d'Italie, 69364 Lyon CEDEX 07, France}
\ead{sylvain.joubaud@ens-lyon.fr, nicolas.garnier@ens-lyon.fr and sergio.ciliberto@ens-lyon.fr}
\begin{abstract}
We study experimentally the thermal fluctuations of energy input and dissipation in a harmonic oscillator driven out of equilibrium, and search for Fluctuation Relations. We study transient evolution from the equilibrium state, together with non equilibrium steady states. Fluctuations Relations are obtained experimentally for both the work and the heat, for the stationary and transient evolutions. A Stationary State Fluctuation Theorem is verified for the two time prescriptions of the torque. But a Transient Fluctuation Theorem is satisfied for the work given to the system but not for the heat dissipated by the system in the case of linear forcing. Experimental observations on the statistical and dynamical properties of the fluctuation of the angle, we derive analytical expressions for the probability density function of the work and the heat. We obtain for the first time an analytic expression of the probability density function of the heat. Agreement between experiments and our modeling is excellent.
\end{abstract}


\newpage
\pagestyle{plain}

\noindent\hrulefill

\tableofcontents

\noindent\hrulefill

\section{Introduction}
Nanotechnology, as well as biology, biophysics and chemistry are using or studying
setups and objects which are smaller and smaller. In these systems, one is usually 
interested in mean values, but thermal fluctuations play an important role because 
their amplitude are often comparable to the mean values. This is for example the
case for quantities such as the energy injected in the system or the energy dissipated 
by the system. These fluctuations can lead to unexpected and undesired effects : for
instance, the instantenous energy transfer can be reversed by a large fluctuation,
leading energy to flow from a cold source to a hot one. These events, although rare,
are quantitatively studied by the recent Fluctuations Theorems (FTs). These theorems 
give Fluctuation Relations (FRs) that quantify the probability of these rare events 
in systems which can be arbitrarily far from equilibrium. FTs have been demonstrated in both deterministic systems~\cite{Evansetal93,GallavottiCohen95} and stochastic dynamics~\cite{Evans-Searles,Kurchan98,Farago,Cohen,Cohen1,Harris06}. 
Experiments searching for FRs have been performed in dynamical systems~\cite{Ciliberto98,Ciliberto04,Menon04}, 
but interpretations are very difficult because a quantitative comparison with theoretical prediction is
impossible. Other experiments have been performed in 
stochastic systems described by a first order Langevin equation: a Brownian particle 
in a moving optical trap~\cite{Wangetal:02:05} and an out-of-equilibrium electrical 
circuit~\cite{Garnier05} in which existing theoretical predictions~\cite{Cohen,Cohen1}
were verified. Interesting comments on the Langevin equation can be found in \cite{Narayan04}. 

In the present article, we study a thermostated harmonic oscillator described by a 
second order Langevin equation. We experimentally search FRs for the work done by an 
external operator and for the heat dissipated by the system, and present analytical 
derivations of FTs based on experimental observations.

This paper is organized as follows. In section \ref{sec:System:description}, we present 
the experimental system, write its energy balance to define the work given to
the system together with the heat dissipated. We then introduce the Fluctuation Relations (FRs) 
and the Fluctuation Theorems (FTs). In sections \ref{sec:TFTexp}, \ref{sec:SSFTrampexp} 
and \ref{sec:sinusexp}, we present experimental results on the fluctuations 
of first the work and then the heat. A short discussion on experimental results 
in given in \ref{sec:expconc}. Then, in sections \ref{sec:worktheo} and \ref{sec:heattheo}, 
we present some analytical derivations of FTs based on hypothesis inspired by experimental 
observations. We compare these analytical predictions to the experimental observations 
and finally conclude in section \ref{sec:conc}.

\section{System description}
\label{sec:System:description}

\subsection{The harmonic oscillator}
\label{sec:pendulum}

\begin{figure}
\centerline{\includegraphics[width=1.0\linewidth]{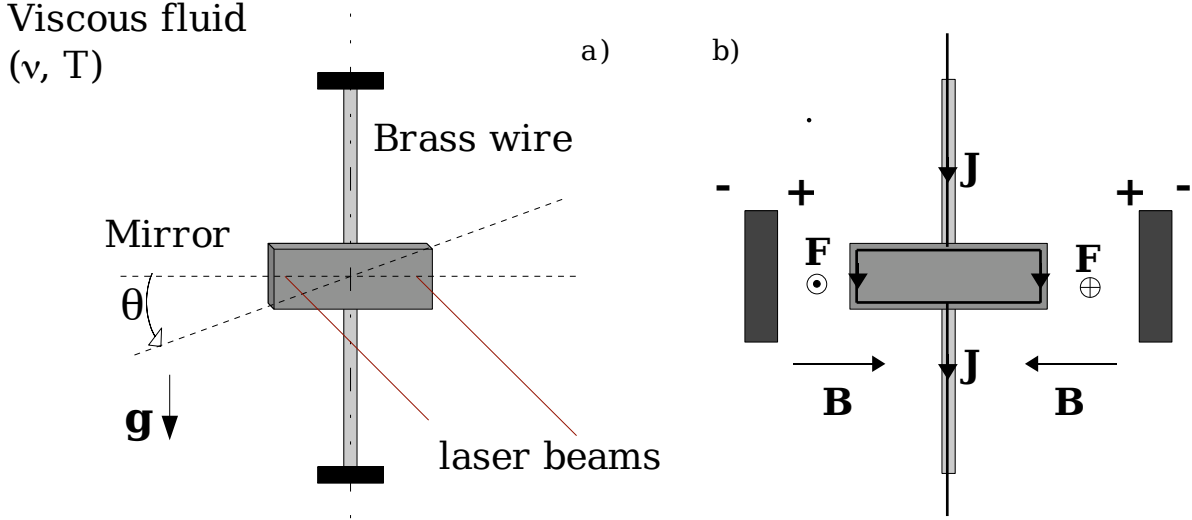}}
\caption{a) The torsion pendulum. b) The magnetostatic forcing.}
\label{fig:pendulum}
\end{figure}

Our system is a harmonic oscillator and we measure the non-equilibrium fluctuations 
of its position degree of freedom. The oscillator is damped due to the viscosity of a 
surrounding fluid that acts as a thermal bath at temperature $T$. Our oscillator, 
depicted in Fig.~\ref{fig:pendulum}a, is a torsion pendulum composed of a brass wire 
(length $10$~$\textrm{mm}$, width $0.5$~$\textrm{mm}$, thickness $50$~$\mu \textrm{m}$) 
and a glass mirror glued in the middle of this wire (length $2$~$\textrm{mm}$, 
width $8$~$\textrm{mm}$, thickness $1$~$\textrm{mm}$). The elastic torsional stiffness 
of the wire is $C = 4.65 \cdot 10^{-4}$ $\textrm{N.m.rad}^{-1}$. It is enclosed in 
a cell filled by a water-glycerol mixture at $60 \%$ concentration. The system is a 
harmonic oscillator with resonant frequency 
$f_o=\sqrt{C/\Ieff}/(2\pi)=\omega_0/(2\pi)=217$~$\textrm{Hz}$ and a relaxation time 
$\ta=2\Ieff/\nu=1/\alpha= 9.5$~$\textrm{ms}$. $\Ieff$ is the total moment of inertia 
of the displaced masses ({\em i.e.} the mirror and the mass of displaced fluid) \cite{Lamb}. 
The damping has two contributions : the viscous damping $\nu$ of the 
surrounding fluid and the viscoelasticity of the brass wire which can be neglected here.

The angular displacement of the pendulum $\theta$ is measured by a differential 
interferometer~\cite{Douarche04, Douarche05,Douarche06}. The measurement noise is two orders of
magnitude smaller than thermal fluctuations of the pendulum. $\theta(t)$ is acquired with a resolution of $24$ bits at a sampling rate of $8192$~$\textrm{Hz}$, which is about 40 times $f_o$. We drive the system out-of-equilibrium by forcing it with an external torque $M$ by means of a small electric current $J$ flowing in a coil glued behind the mirror (Fig.~\ref{fig:pendulum}b). The coil is inside a static magnetic field. The displacements of the coil and therefore the angular displacements of the mirror are much smaller than the spatial scale of inhomogeneity of the magnetic field. So the torque is proportional to the injected current : $M = A.J$ ; the slope $A$ depends on the geometry of the system.

The angular displacement $\theta$ of this harmonic oscillator is very well described by a second order Langevin equation:
\begin{equation}
\Ieff\,\frac{\dd^2{\theta}}{\dd t^2}+ \nu \,\frac{\dd{\theta}}{\dd t}  + C\,\theta =
M + \sqrt{2k_B T\nu} \, \eta,
\label{eq:Langevin_oscillator}
\end{equation}
where $\eta$ is the thermal noise, delta-correlated in time of variance $1$ and $k_B$ the Boltzmann constant and $T$ the temperature of the system which is the one 
of the surrounding fluid. The Fluctuation Dissipation Theorem (FDT) gives a relation 
between the amplitude of the thermal angular fluctuations of the oscillator at 
equilibrium and its response function. For a harmonic oscillator, the equilibrium 
thermal fluctuation power spectral density (psd) is:
\begin{equation}
\langle |\hat{\theta}|^2\rangle \, = \, \frac{4 k_B T}{\omega}\,\textrm{Im} \hat \chi =
\frac{4 k_B T \nu}{(-\Ieff \omega^2+C)^2+(\omega \nu)^2}
\label{eq:FDT}
\end{equation}
where $\hat \chi = \frac{\hat M}{\hat \theta} = A \frac{\hat J}{\hat \theta}$. Using 
FDT (Eq.~\ref{eq:FDT}), we measure the coefficient $A$ and test the calibration accuracy 
of the apparatus which is better than $3\%$. More details on the set-up can be found 
in~\cite{Douarche04, Douarche05}.

\subsection{Energy balance}

When the system is driven out of equilibrium using a deterministic torque, it receives some work and a fraction of this energy is dissipated into the heat bath. Multiplying Eq.~(\ref{eq:Langevin_oscillator}) by $\dot{\theta}$ and integrating between $t_i$ and $t_i+\tau$, we obtain a formulation of the first law of thermodynamics between the two states at time $t_i$ and $t_i+\tau$ (Eq.~(\ref{eq:Energy_conservation})). The change in internal energy $\DUt$ of the oscillator over a time $\tau$, starting at a time $t_i$, is written as:
\begin{equation}
\DUt = U(t_i+\tau) - U(t_i) = \Qt + \Wt
\label{eq:Energy_conservation}
\end{equation}
where $\Wt$ is the work done on the system over a time $\tau$ :
\begin{equation}
\Wt = \frac{1}{k_B T} \int_{t_i}^{t_i+\tau} M(t') \frac{\dd\theta}{\dd t}(t') \dd t' 
\label{eq:Wdef}
\end{equation}
and $\Qt$ is the heat given to the system. Equivalently, $(-\Qt)$ is the heat dissipated by the system. $\DUt$, $\Wt$ and $\Qt$ are defined as energy in $k_B T$ units. The internal energy is the sum of the potential energy and the kinetic energy :
\begin{equation}
U(t)=\frac{1}{k_B T} \left\{\frac{1}{2} \Ieff\left[
\frac{\dd \theta}{\dd t}(t) \right]^{2} +\frac{1}{2} C
\theta(t)^2 \right\}.
\label{eq:Udef}
\end{equation}
The heat transfer $\Qt$ is deduced from equation (\ref{eq:Energy_conservation}) ; it has two contributions :
\begin{eqnarray}
\Qt &=& \DUt - \Wt \nonumber\\
&=& -\frac{1}{k_B \ T }
\int_{t_i}^{t_i+\tau} \nu \left[ \frac{\dd \theta}{\dd t}
(t')\right]^{2}\dd t' +\frac{1}{k_B \ T } \int_{t_i}^{t_i+\tau}
\eta(t') \frac{\dd \theta}{\dd t}(t') \dd t' .
\label{eq:Qdef}
\end{eqnarray}
The first term corresponds to the opposite of viscous dissipation and is always negative, whereas the second term can be interpreted as the work of the thermal noise which have a fluctuating sign. The second law of thermodynamics imposes $\langle -\Qt \rangle$ to be positive. We rescale the work $\Wt$ (the heat $\Qt$) by the average work $\langle \Wt \rangle$ (the average heat $\langle \Qt \rangle$) and define: $\wt = \frac{\Wt}{\langle \Wt \rangle}$ ($\qt = \frac{\Qt}{\langle \Qt \rangle}$). The brackets are ensemble averages. In the present article, $x_\tau$, respectively $X_\tau$, stands for either $\wt$ or $\qt$, respectively $\Wt$ or $\Qt$.

\subsection{Fluctuation Theorems and Fluctuation relations}
There are two classes of FTs. The {\it Stationary State Fluctuation Theorem} (SSFT) considers a non-equilibrium steady state. The {\it Transient Fluctuation Theorem} (TFT) describes transient non-equilibrium states where $\tau$ measures the time since the system left the equilibrium state.
A Fluctuation Relation (FR) examines the symmetry of the probability density function (PDF) $p(x_\tau)$ of a quantity $x_\tau$ around $0$ ; $x_\tau$ is an average value over a time $\tau$. It compares the probability to have a positive event ($x_\tau = +x$) versus the probability to have a negative event ($x_\tau = -x$). We quantify the FR using a function $S$ (symmetry function) :
\begin{equation}
S(x_\tau) = \frac{1}{\langle X_\tau \rangle}\ln \left( \frac{p(x_\tau=+x)}{p(x_\tau=-x)}\right).
\label{eq:FT}
\end{equation}

The {\it Transient Fluctuation Theorem} (TFT) states that the symmetry function is linear with $x_\tau$ for any values of the time integration $\tau$ and  the proportionality coefficient is equal to $1$ for any value of $\tau$.
\begin{equation}
S(x_\tau)=x_\tau, \quad \forall x_\tau, \quad \forall \tau.
\label{eq:TFT}
\end{equation}
Contrary to TFT, the {\it Stationary State Fluctuation Theorem} (SSFT) holds only in the limit of infinite time ($\tau$).
\begin{equation}
\lim_{\tau \rightarrow \infty} S(x_\tau) = x_\tau, \quad \forall x_\tau.
\label{eq:SSFT}
\end{equation}

The questions we ask are whether fluctuation relations for finite time satisfy the two theorems and what are the finite time corrections. In a first time, we test the correction to the proportionality between the symmetry function $S(x_\tau)$ and $x_\tau$. In the region where the symmetry function is linear with $x_\tau$, we define the slope $\Sigma_x(\tau)$ : $S(x_\tau) = \Sigma_x(\tau) x_\tau$. In a second time we measure finite time corrections to the value $\Sigma_x(\tau) = 1$ which is the asymptotic value expected by the two theorems.

\section{Transient non-equilibrium state\label{sec:TFTexp}}
\begin{figure}
\centerline{\includegraphics[width=1.0\linewidth]{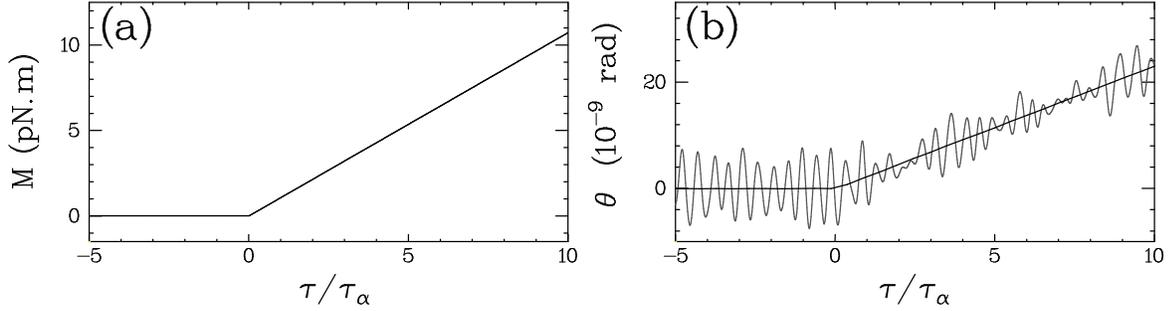}}
\caption{ a) Typical driving torque applied to the oscillator; b)
 Response of the oscillator to the external torque (gray line).
 The dark line represents the mean response $\bar \theta(t)$
 to the applied torque $M(t)$.}
\label{fig:lineartorque}
\end{figure}
For the Transient Fluctuation Theorem, we choose the torque $M(t)$ depicted in Fig.~\ref{fig:lineartorque}a). It is a linear function of time : $M(t)\,=\,M_o\,t/\tau_r$ with $M_o\,=\,11.28$~$\textrm{pN.m}$ and $\tau_r\,=\,0.1$~$\textrm{s}\, = 10.52\,\ta$. The value of $M_o$ is chosen such that the mean response of the oscillator is of order of the thermal noise, as can be seen in Fig.~\ref{fig:lineartorque}b) where $\theta(t)$ is plotted during the same time interval of Fig.~\ref{fig:lineartorque}a). The system is at equilibrium at $t_i = 0$ ($M(t_i=0) = 0$ pN.m and $M(t) = 0$ pN.m $\forall t<t_i$). In this section the starting time $t_i$ of integration of all quantities defined before ($\Wt$, $\DUt$ and $\Qt$) is $t_i = 0$. So the work is :
\begin{equation}
\Wt = \frac{1}{k_B \ T } \int_{0}^{\tau} M(t') \frac{\textrm d \theta}{\textrm dt}(t') dt'.
\label{eq:def_Work_ramps}
\end{equation}
\subsection{Average values informations}
\begin{figure}
\centerline{\includegraphics[width=1.0\linewidth]{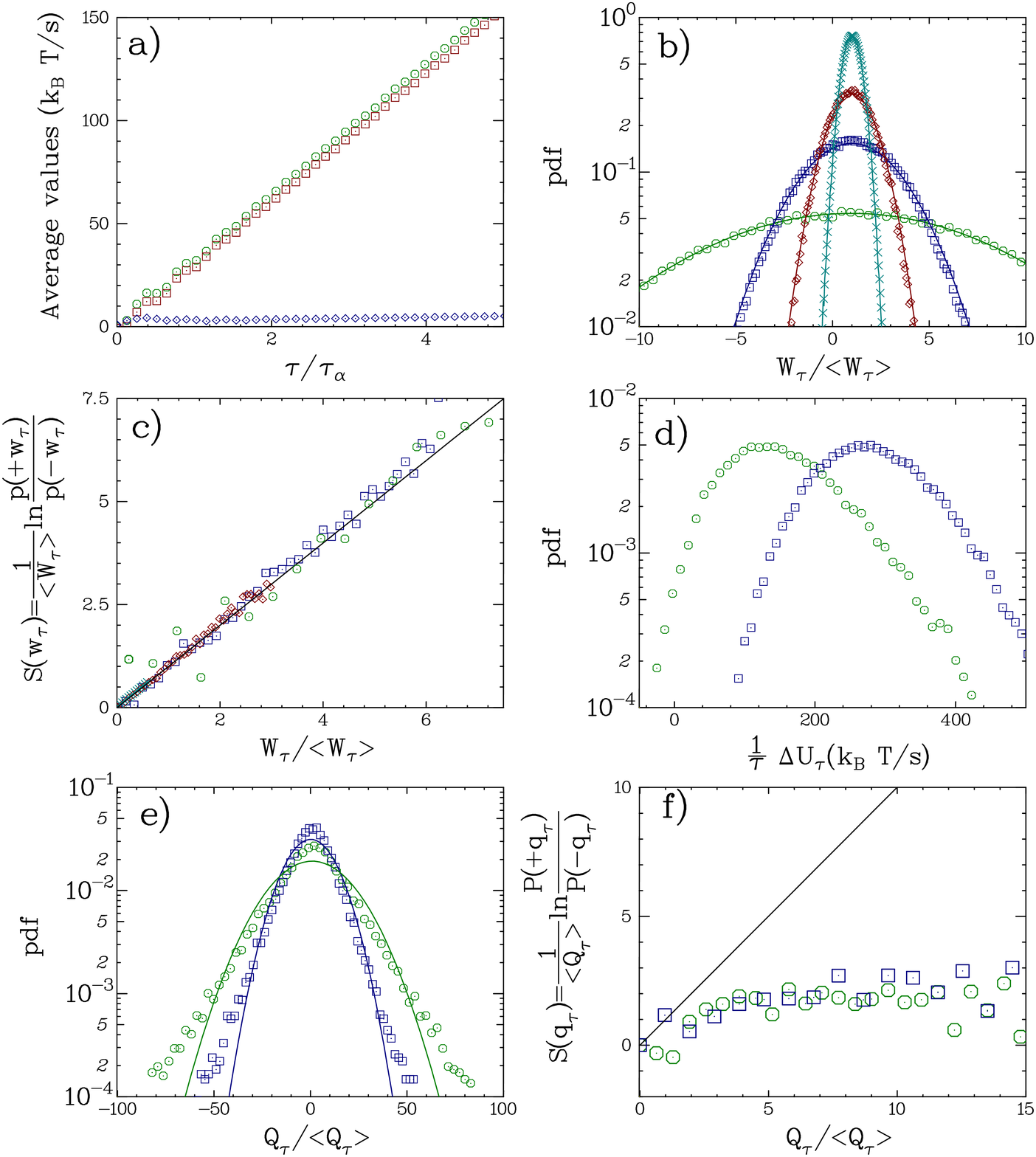}}
\caption{TFT. a) Average value of $\tau^{-1} \Wt$ $(\circ)$, $\tau^{-1} \DUt$ $(\Box)$ and $\tau^{-1} \Qt$ $(\diamond)$ plotted as a function of $\tau$. b) PDFs of $\wt$ for various $\tau/\ta$ : 0.31 $(\circ)$, 1.015 $(\Box)$, 2.09 $(\diamond)$ and 4.97 $(\times)$. Continuous lines are theoretical predictions with no adjustable parameters. c) Corresponding functions $S(w_\tau)$. The straight continuous line is a line with slope $1$. d) PDFs of $\tau^{-1} \DUt$ for two values of $\tau/\ta$ : 4.97 $(\circ)$ and 8,96 $(\Box)$. e) Corresponding PDFs of $\qt$. Continuous lines are Gaussian fits. f) Corresponding functions $S(q_\tau)$. The straight continuous line is a line with slope $1$.}
\label{fig:TFT}
\end{figure}

In Fig.~\ref{fig:TFT}a), we represent the time average ($\langle\tau^{-1} \Wt\rangle$) of the power injected into the system, the internal energy difference $\langle \tau^{-1} \DUt \rangle$ and the time average  ($\langle\tau^{-1} \Wt\rangle$) of the power dissipated by the system. $\langle\tau^{-1} \Wt\rangle$ and $\langle \tau^{-1} \DUt \rangle$ are linear in $\tau$ after some short relaxation time $\ta$ defined in the Langevin equation : for $\tau/\ta$ smaller than $1$, some oscillations around the linear behavior can be seen. The average value of work $\langle \Wt \rangle$ is therefore quadratic in $\tau$ and is equal to $33$~$k_B T$ for $\tau=\tau_r$. The difference between $\langle \Wt \rangle$ and $\langle \DUt \rangle$ corresponds to the mean value of dissipated heat $\langle -\Qt  \rangle$ (Eq.~(\ref{eq:Qdef})). As can be seen in fig.\ref{fig:TFT}a), $\langle \Wt \rangle $ is larger than $\langle \DUt \rangle$ for all times $\tau$. The average of the dissipated power ($\langle - \tau^{-1}\Qt \rangle$) is therefore positive for all times $\tau$ as expected from the second principle. For $\tau$ larger than several $\ta$, the dissipated power is constant and equal to a few $k_B T$ per second because $\langle \tau^{-1} \Wt \rangle$ and $\langle \tau^{-1} \DUt \rangle$ have the same slope after some $\ta$. So we have the following behavior : the work done by the external work is used by the system to increase its internal energy but a small amount of energy is lost at a constant rate by viscous dissipation and exchange with thermostat.

\subsection{Work fluctuations}
The probability density functions (PDFs) $p(\Wt)$ of $\wt$ is plotted in Fig.~\ref{fig:TFT}b) for different values of $\tau/\ta$.  Four typical value of $\tau$ are presented : the first is smaller than the relaxation time and the last equals five relaxation times ; the results are the same for any value of $\tau$. The PDFs  of $\wt$ are Gaussian for any $\tau$. We observe that $\wt$ takes negative values as long as $\tau$ is not too large. The probability of having negative values of $\wt$ decreases when $\tau$ is increased. From the PDFs, we compute the symmetry functions. They are plotted in Fig.~\ref{fig:TFT}c) as a function of $\wt$. They all collapse on the same linear function of $\wt$ for any $\tau$, which implies they all have the same slope $\Sigma_w(\tau)$. The straight line in fig~\ref{fig:TFT}c) has slope $1$. Within experimental error bars, $\Sigma_w(\tau)$ is equal to $1$ for all time $\tau$. Therefore work fluctuations for a harmonic oscillator under a linear forcing satisfy the TFT. We checked that this property is true for other values of $M_o$ and $\tau_r$.

\subsection{Heat fluctuations\label{sec:heatTFT}}

The PDFs of $\tau^{-1} \DUt$ are plotted in Fig.~\ref{fig:TFT}d) for two values of $\tau/\ta$ : they are not symmetric and have exponential tails. The PDFs of $\qt$ can be seen in Fig.~\ref{fig:TFT}e) for the same values of $\tau/\ta$. They are qualitatively different from the ones of the work. We have plotted in the same figure the Gaussian fit of the two PDFs of the dissipated heat. It is clear that the PDFs of $\qt$ are not Gaussian. Extreme events of $\qt$ are distributed on exponential tails. These tails can be interpreted noticing that $\Qt = \DUt-\Wt$ and $\DUt$ have exponential tails. The variance of the PDFs of $\qt$ is also much larger than the variance of the PDFs of $\wt$.

We plot on Fig.~\ref{fig:TFT}f) symmetry functions $S(\qt)$ for the same times $\tau/\ta$. Only the behavior of large events can be analyzed here because the variance is much larger than the mean $\sigma_{\wt} \gg 1$.  As it can be seen in Fig.~\ref{fig:TFT}f, $S(\qt)$ is not proportional to  $\qt$, therefore TFR is not satisfied for finite time. Within experimental resolution, $S(\qt)$ is constant for extreme events and equal to $2$. This behavior can be interpreted by writing for large $\qt$, $p(\qt) = A_{\pm} \exp (-\alpha_{\pm} |\qt|)$ where $\alpha_+$ and $\alpha_-$ are the decrease rate on the exponential tails. Each coefficient depends on $\tau$. 
There is a simple expression of $S(\qt)$ for large fluctuations:
\begin{equation}
S(\Qt) = (\alpha_+ - \alpha_-) \Qt + \frac{1}{\langle \Qt \rangle} \ln \left( \frac{A_+}{A_-} \right).
\label{eq:TFTlargefluctuation}
\end{equation}
In Fig.~\ref{fig:TFT}c), the PDFs of $\qt$ are symmetric around the mean value  for the two values of $\tau$. It is not the case for small $\tau/\ta$. Thus we can conclude that $\alpha_+ = \alpha_-$ and that the symmetry function is so equal to the constant : $(\langle \Qt \rangle)^{-1} (\ln(A_+)-\ln(A_-))$.

As it can be seen in Fig.~\ref{fig:TFT}e), the PDFs become more and more Gaussian when $\tau$ tends to infinity. It is expected that for infinite time, the PDF of $\qt$ is a Gaussian. Thus, TFT appears to be satisfied experimentally in the limit of infinite $\tau$. Our interesting finding is that, for $\Qt$ TFT if not valid for any times.

\section{Steady state : linear forcing\label{sec:SSFTrampexp}}
\subsection{Definition of the work given to the system}
We call a steady state a state in which both forcing and response to the forcing do not depend on the initial time $t_i$, but only on $\tau$. This implies that $\langle M(t_i+\tau) \rangle$ i ndependant of $t_i$ ; and so is $\langle \theta(t_i+\tau)\rangle$.
If the torque drifts along time, the mean of $M(t_i+\tau)$ is linear with $t_i+\tau$. Thus we have to change the definition of the work done on the system to be in a steady state. This is equivalent to a Galilean change of reference frame. The work is now defined as :
\begin{equation}
\Wt = \frac{1}{k_B T}\int_{t_i}^{t_i+\tau} [M(t)-M(t_i)] \frac{\textrm d \theta}{\textrm dt}(t') dt'.
\label{eq:work_SSFT_def}
\end{equation}
With this definition, the forcing is $M(t)-M(t_i)$ and the response to the forcing $\theta(t)-\theta(t_i)$.
When we impose a forcing linear in time ($M(t)=M_o\,t/\tau_r$), the first condition ($\langle M(t_i+\tau)\rangle$ independent of $t_i$) is satisfied. The second ($\langle \theta(t_i+\tau)\rangle$ independent of $t_i$) is also satisfied if $t_i\geq3\ta$, {\em{i.e.}} after a transient state. Thus the system is on a steady state. We remark that, in the transient state, this definition of the work reduces to the usual one, because $M(t_i)=M(t=0)=0$~$\textrm{pN.m}$.

\subsection{Work fluctuations}
\begin{figure}
\centerline{\includegraphics[width=1.0\linewidth]{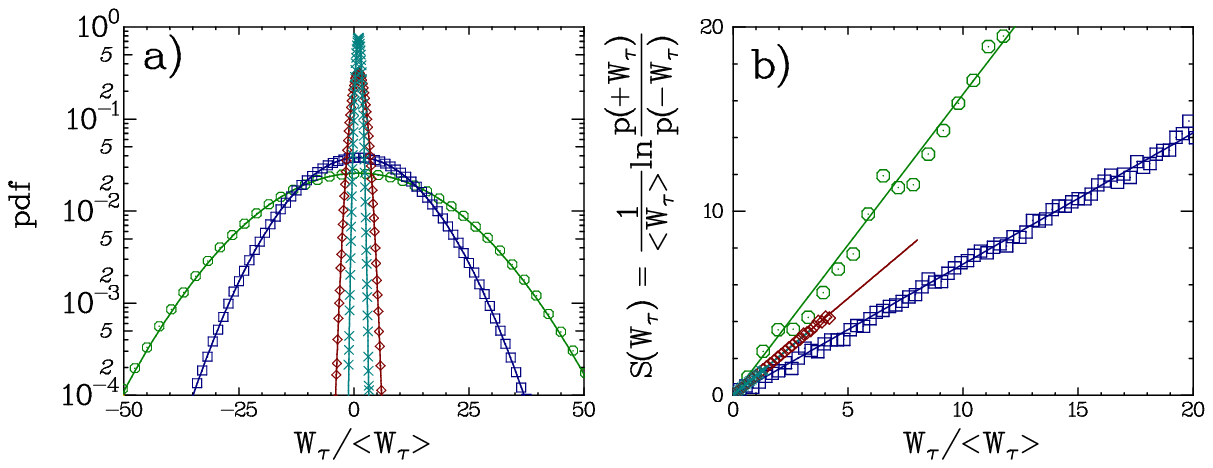}}
\centerline{\includegraphics[width=0.8\linewidth]{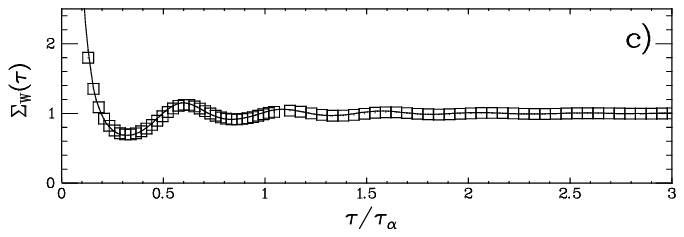}}
\caption{SSFT with a ramp forcing. 
a) PDF of $\Wt$ for various $\tau/\ta$: 0.019 $(\circ)$, 
0.31 $(\Box)$, 2.09 $(\diamond)$ and 4.97 $(\times)$. 
b) Corresponding functions $S(\Wt)$. 
c) The slope $\Sigma_w(\tau)$ of $S(\Wt)$ is plotted
versus $\tau$ ($\Box$: experimental values; continuous line: 
theoretical prediction eq.(\ref{eq:epsilon_linear}) with no adjustable parameters).}
\label{fig:SSFTramp}
\end{figure}

The average of $\Wt$ is quadratic in $\tau$ for any value of $\tau/\ta$. There are no oscillations in time for small $\tau/\ta$. The PDFs of $\wt$ are Gaussian for any value of $\tau/\ta$ (Fig.~\ref{fig:SSFTramp}a). Probability of negative values is high and decreases with $\tau$, like in the transient case. The symmetry functions $S(\wt)$ are again proportional to $\wt$ (Fig.~\ref{fig:SSFTramp}b) but the slope $\Sigma_w$ is not equal to $1$ for smaller $\tau$ and tends to $1$ for $\tau \gg \ta$ only, as can be seen in Fig.~\ref{fig:SSFTramp}c. Thus we obtain a fluctuation relation for the work done on the system in this steady state and this relation satisfy the SSFT. The slope at finite time is slightly oscillating at a frequency, close to $f_0$.

\subsection{Heat fluctuations}
The heat dissipated during this linear forcing has a behavior very similar to the one observed in the transient case (section \ref{sec:heatTFT}). We can so transpose here all what we said in section \ref{sec:heatTFT}.

\section{Steady state : sinusoidal forcing\label{sec:sinusexp}}
\begin{figure}
\centerline{\includegraphics[width=1.0\linewidth]{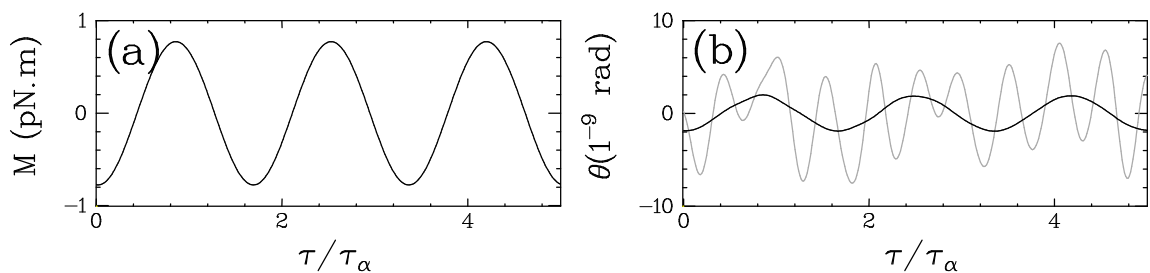}}
\caption{a) Sinusoidal driving torque applied to the oscillator. b) Response of the oscillator to this periodic forcing (gray line) ; the dark line represents the mean response $\langle \theta(t) \rangle$.}
\label{fig:sinustorque}
\end{figure}

We now consider a periodic forcing $M(t) = M_o \sin(\omega_d t)$. This is a very common kind of forcing but it has never been studied in this context. Using Fourier transform, any periodical forcing can be decomposed in a sum of sinusoidal forcing. We explain here the behavior of a single mode. We choose $M_o = 0,78$~$\textrm{pN.m}$ and $\omega_d/(2\pi) = 64$~$\textrm{Hz}$. This torque is plotted in Fig.~\ref{fig:sinustorque}a. The mean of the response to this torque is sinusoidal, with the same frequency, as can be seen in Fig.~\ref{fig:sinustorque}b. We studied other frequencies $\omega_d$. The system is clearly in a steady state. We choose the integration time $\tau$ to be a multiple of the period of the driving ($\tau = 2n\pi/\omega_d$ with $n$ integer). The starting phase $t_i\omega_d$ is averaged over all possible $t_i$ in order to increase statistics ; in the remaining of this section, we drop the brackets $\langle .\rangle_{t_i}$.

\subsection{Work fluctuations}
\begin{figure}
\centerline{\includegraphics[width=1.0\linewidth]{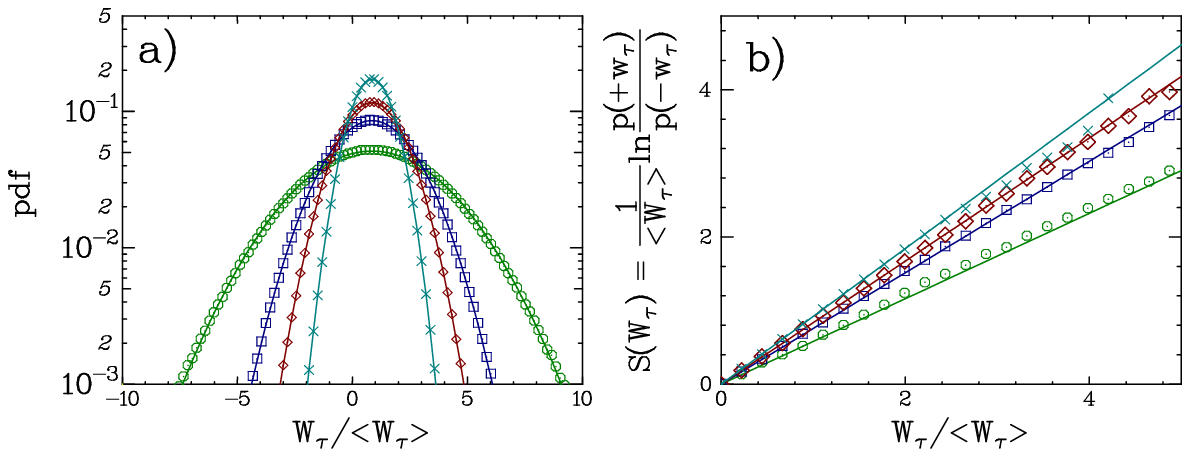}}
\centerline{\includegraphics[width=0.8\linewidth]{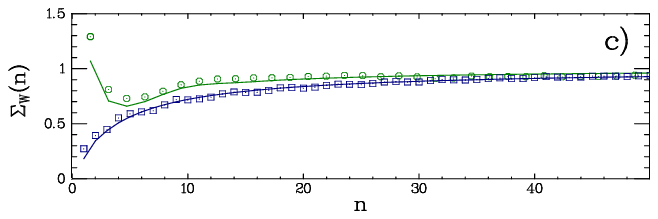}}
\caption{Sinusoidal forcing.
a) PDFs of the work $\wn$ integrated over $n$ periods of forcing, with
$n=7$ ($\circ$), $n=15$ ($\Box$), $n=25$ ($\diamond$) and $n=50$ ($\times$). 
b) The function $S(\wn)$ measured at $\omega_d/2\pi=64$~$\textrm{Hz}$ is plotted
as a function of $\wn$ for several $n$: $(\circ) n=7$;
$(\Box) n=15$ $(\diamond) n=25$; $(\times) n=50$. For these two plots, continuous lines are theoretical predictions with no adjustable paramaters (Eq.~(\ref{eq:SSFT_sinus_mean_PDF_work}) and Eq.~(\ref{eq:SSFT_sinus_Var_PDF_work})).
c) The slopes $\Sigma_w(n)$, plotted as a function of $n$ for two different driving frequencies $\omega_d$ = 64 Hz ($\Box$) and 256 Hz ($\circ$); continuous lines are theoretical predictions from Eq.~(\ref{eq:epsilon_sinus}) with no adjustable parameters.} 
\label{fig:sinuswork}
\end{figure}

The work is written as a function of $n$, the number of periods of the forcing :
\begin{equation}
\Wn =\frac{1}{k_B T} \int_{t_i}^{t_i+\tau_n} M(t')\frac{\dd \theta}{\dd t}(t') \dd t'.
\label{eq:worksinusdef}
\end{equation}
The PDFs of $\wn$ are plotted in Fig.~\ref{fig:sinuswork}a. Work fluctuations are Gaussian for all values of $n$ as in previous cases. Thus symmetry functions are again linear in $\wn$ (Fig.~\ref{fig:sinuswork}b). The slope $\Sigma_w(n)$ is not equal to $1$ for all $n$ but there is a correction at finite time (Fig.~\ref{fig:sinuswork}c). Nevertheless, $\Sigma_w(n)$ tends to $1$ for large $n$, so SSFT is satisfied. The convergence is very slow and we have to wait a large number of periods of forcing for the slope to be $1$ (after $30$ periods, the slope is still $0.9$).

This behavior is independent of the amplitude of the forcing $M_o$ and consequently of the mean value of the work $\langle W_n \rangle$. The system satisfies the SSFT for all forcing frequencies $\omega_d$ but finite time corrections depends on $\omega_d$, as can be seen in Fig.~\ref{fig:sinuswork}c.
\subsection{Heat fluctuations}
\begin{figure}
\centerline{\includegraphics[width=\linewidth]{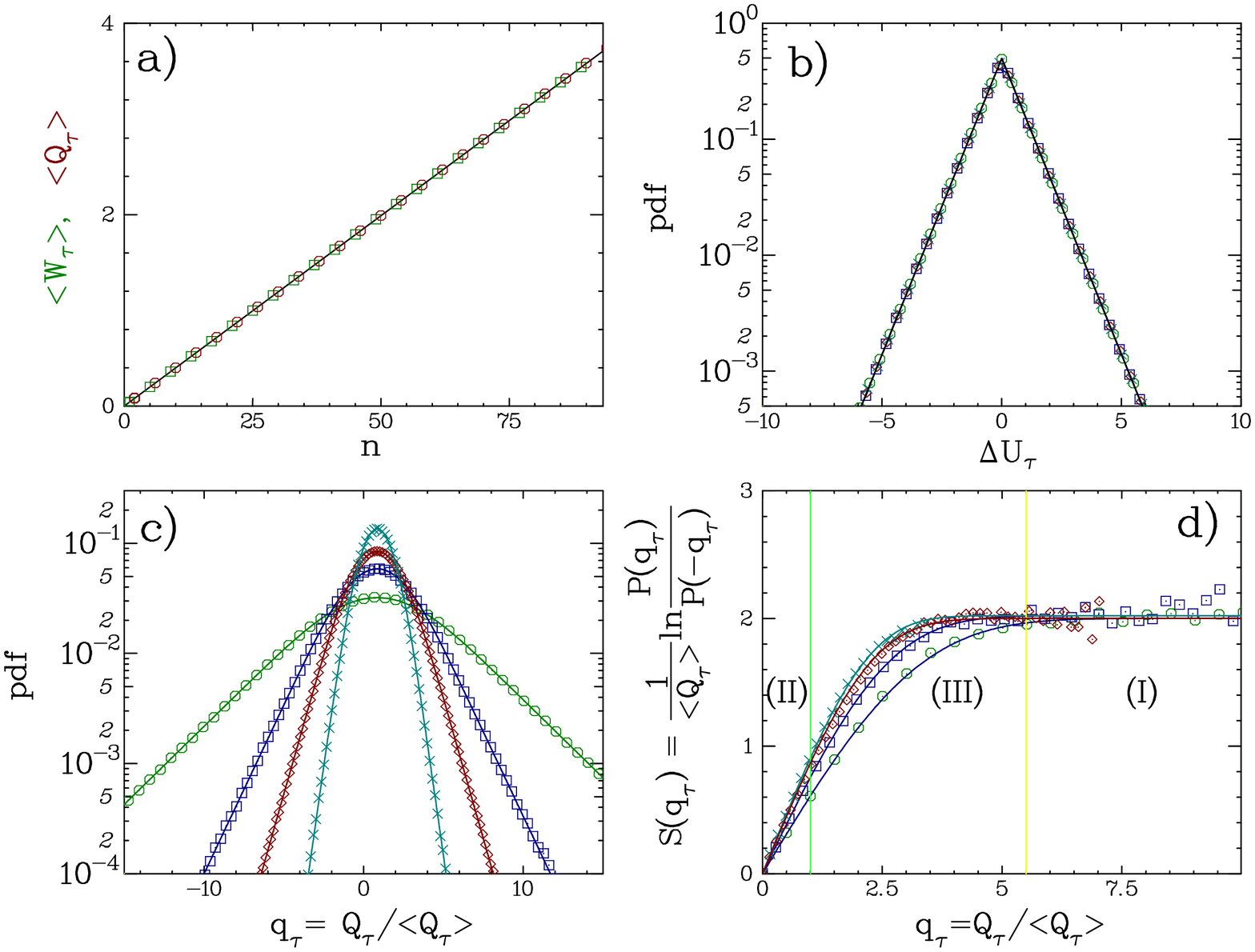}}
\centerline{\includegraphics[width=0.7\linewidth]{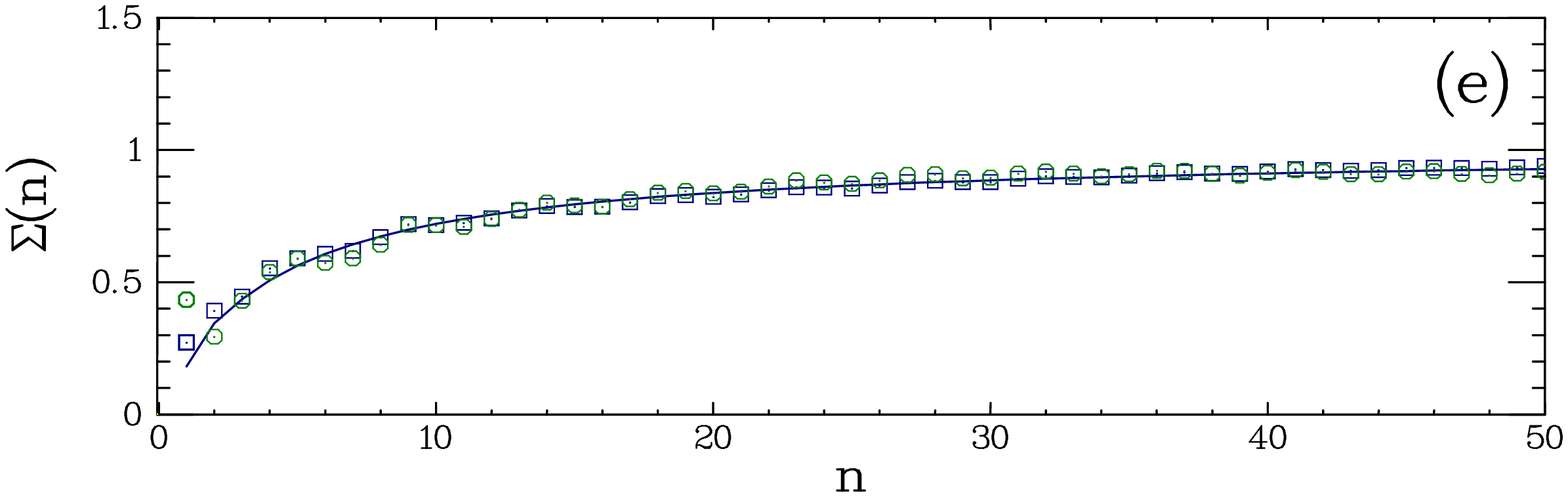}}
\caption{Sinusoidal forcing. a) Average value of $\Wn$ ($\circ$) and $\Qn$ ($\Box$).  In the next plots, the integration time $\tau$ is a multiple of the period of forcing, $\tau = 2 n \pi/\omega_d$, with $n=7$ ($\circ$), $n=15$ ($\Box$), $n=25$ ($\diamond$) and $n=50$ ($\times$). Continuous lines are theoretical predictions with no adjustable parameters.  b) PDFs of $\DUt$. c) PDFs of $\qt$. d) Symmetry functions $S(\qt)$. e) The slope $\Sigma_q(n)$ of $S(\qt)$ for $\qt <1$, plotted as a function of $n$ ($\circ$). The slope $\Sigma_w(n)$ of $S(\wt)$ plotted as a function of $n$($\Box$). Continuous line is theoretical prediction.}
\label{fig:sinusheat}
\end{figure}

We first do some comments on the average values. The average of $\DUt$ is obviously vanishing because the time $\tau$ is a multiple of the period of the forcing. $\langle \wn \rangle$ and $\langle \qn \rangle$ have consequently the same behavior and they are linear in $\tau$, as can be seen in Fig.~\ref{fig:sinusheat}a) but the PDFs of heat fluctuations $\qn$ have exponential tails (Fig.~\ref{fig:sinusheat}c). This can be understood noticing that, from Eq.~(\ref{eq:Qdef}),
$-\Qt=\Wt-\DUt$ and that $\DUt$ has an exponential PDF independent of $n$ (Fig.~\ref{fig:sinusheat}b). Therefore, in a first approximation, the PDF of $\qt$ is a convolution between an exponential distribution (PDF of $\DUt$) and a Gaussian distribution (PDF of $\wt$).

Symmetry functions $S(\qn)$ are plotted in Fig.~\ref{fig:sinusheat}d) for different values of $n$ ; three different regions appear:

(I) For large fluctuations $\qn$, $S(\qn)$ equals $2$.
When $\tau$ tends to infinity, this region spans from $\qn = 3$ to infinity.

(II) For small fluctuations $\qn$, $S(\qn)$ is a linear function of $\qn$.
We then define $\Sigma_q(n)$ as the slope of the function $S(\qn)$, {\it i.e.} $S(\qn) = \Sigma_q(n) \, \qn$. This slope is plotted in Fig.~\ref{fig:sinusheat}e) where we see that it tends to $1$ when $\tau$ is increased.
So, SSFT holds in this region II which spans from $\qn=0$ up to $\qn = 1$ for large $\tau$.

(III) A smooth connection between the two behaviors.

We observe that $\Sigma_w(n)$ matches experimentally $\Sigma_q(n)$, for all values of $n$ (Fig.~\ref{fig:sinusheat}e). So the finite time corrections to the FT for the heat are the same than the ones of FT for work : $\Sigma_w(n)=\Sigma_q(n)$.

These regions define the Fluctuation Relation from the heat dissipated by the oscillator. The limit for large $\tau$ of the symmetry function $S(\qt)$ is rather delicate and we will discuss it in section \ref{sec:heat_sinus_theo}.

\section{Discussion and conclusion on experimental results}
\label{sec:expconc}

In the previous sections, we have presented experimental results on a 
harmonic oscillator driven out of equilibrium by an external deterministic forcing $M$. 
We operated with two different time-prescriptions : one in which $M$ is a linear function of
time, and one in which $M$ is a sinusoidal function of time.

The energy injected into the system is the work $W$ of the torque $M$.
The PDFs of the work $W$ are Gaussian whatever the time prescription of $M$ is, and work 
fluctuations satisfy a TFT ($M$ linear in time) and a SSFT ($M$ linear or sinusoidal in time). 

The energy dissipated by the system is represented by the heat $Q$, and we measured it
using the first principle of thermodynamics (eq.~\ref{eq:Qdef}).
Heat probability distributions are not Gaussian and are very different from the ones of the work. 
They nevertheless satisfy a SSFT in both the case of a sinusoidal forcing and a linear forcing. 
But they do not satisfy a TFT in the case of a linear forcing, because the symmetry functions 
are not linear for all values of dissipated heat $\qt$. 

In the next two sections, we use some experimental evidences to derive 
analytical expressions of the PDFs of work and the heat exchanged on an arbitrary time interval $\tau$. 
We then derive FTs together with their finite time corrections.

\section{Work fluctuations : theoretical predictions\label{sec:worktheo}}

In this section, we derive the analytical expression of the PDF of the work 
given to the system, and defined as the work of the torque applied to the pendulum,
which is either linear or sinusoidal in time.
Experimentally, we observed that the PDFs are always Gaussian, so we restrict our task to deriving 
expressions for the first two moments of the work distribution.

To do so, we use experimental observations on the fluctuations of the angle $\theta$, as described 
in section~\ref{sec:exp_observations} below. We then compute in section \ref{sec:pdf_W} the mean and the variance of 
the work $W_\tau$ in the different experimental situations, and then 
write formally the corresponding Fluctuations Relations, from which we obtain analytical 
expressions of the finite time corrections to the Fluctuation Theorems.

\subsection{Angular fluctuations in the presence of forcing.}
\label{sec:exp_observations}
We discuss here the angular fluctuations. We decompose the angle $\theta$ into a mean value $\langle \theta \rangle$ 
and a fluctuating part $\delta \theta$, writing $\theta = \langle \theta \rangle + \delta \theta$. The mean value
corresponds to an ensemble average. It is obtained experimentally by averaging over realisations of the forcing,
and it is presented in Fig.~\ref{fig:lineartorque} and \ref{fig:sinustorque}. 

A first experimental observation is as follows. The measured mean response $\langle \theta \rangle$ 
is exactly equal to the solution of the deterministic second order equation obtained when removing 
the noise term ($\eta = 0$) in the Langevin equation~(\ref{eq:Langevin_oscillator}). We checked this
from our data, and found this way a value of the calibration $A$ (see section~\ref{sec:pendulum})
in perfect agreement with the one obtained from the application of the Fluctuation Dissipation Theorem.

\begin{figure}
\centerline{\includegraphics[width=1.0\linewidth]{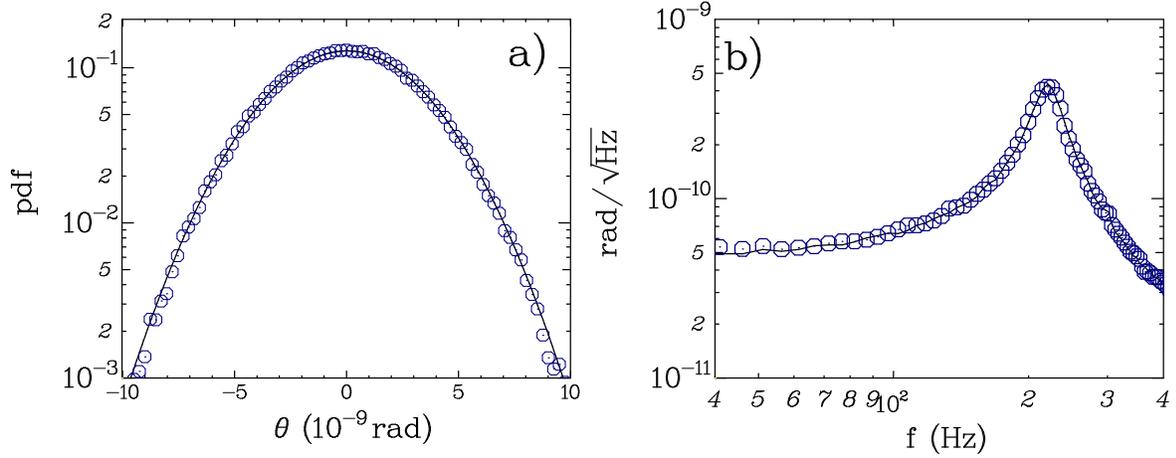}}
\caption{a) PDF of the fluctuations $\delta
\theta= \theta -\bar \theta(t)$ when the torque is applied ($\circ$), 
compared with a Gaussian fit of the PDF at equilibrium (continuous line). 
b) The measured spectrum of $\delta \theta$ ($\circ$) is compared with the prediction
of fluctuation dissipation theorem in equilibrium (continuous line).}
\label{fig:fluct_ang_comparison}
\end{figure}

A second experimental observation concerns the probability distribution of $\delta \theta$
in out-of-equilibrium conditions. We know and observed that at equilibrium, $\delta \theta$ 
has a Gaussian distribution with variance $\sigma_\theta^2 = k_B\, T/C$, 
and the associated momentum $\dot{\theta}$ has fluctuations $\delta \dot{\theta}$ 
which also have a Gaussian distribution, with a variance $k_B \,T/ \Ieff$. 
We observe that the statistical properties of angular fluctuations 
$\delta \theta$ when a torque $M(t)$ linear in time is applied are the same 
as the statistical properties at equilibrium, when no torque is applied. 
In figure~\ref{fig:fluct_ang_comparison}a, we plot the PDF of $\delta \theta$ 
measured at $M\neq0$ together with the Gaussian fit of the PDF at equilibrium (continuous line). 
The two curves matches perfectly within experimental accuracy. 
Thus we conclude that the external driving does not perturb the 
equilibrium distribution of angular fluctuations, so we use:
\begin{equation}
P(\delta \theta,M\neq0)=P(\delta \theta, M=0) = \frac{1}{\sqrt{2\pi\sigma_\theta^2}}\exp{\left(-\frac{\delta\theta^2}{2\sigma_\theta^2}\right)}.
\label{PDFtheta}
\end{equation}
%
%

The third experimental observation concerns time correlations.
In figure~\ref{fig:fluct_ang_comparison}b, we plot the power spectral density function 
of $\delta \theta$ when applying an external forcing ($\circ$). We compare it to the prediction 
of fluctuation dissipation theorem at equilibrium (Eq.~\ref{eq:FDT}) computed using the 
oscillator parameters. The two spectra are identical, so we can confidently
use for our system a description in terms of a second order Langevin dynamic where the noise term is not perturbed by the presence of the driving.
From the power spectral density function of $\theta$ (Eq.~\ref{eq:FDT}), we derive the autocorrelation function $R_{\delta \theta}(\tau)$ of $\delta \theta$ 
during a time interval $\tau$. It is the same at equilibrium and out of equilibrium, and
decreases exponentially:
\begin{equation}
R_{\delta \theta}(\tau) = \langle \delta \theta(t+\tau) \delta \theta(t)\rangle 
= \frac{k_B\,T}{C\, \sin(\varphi)}\,\exp\left(-\frac{|\tau|}{\ta}\right) \sin(\psi |\tau| + \varphi)
\label{eq:autocorrelation}
\end{equation}
where $\psi^2 \equiv (\omega_0)^2-(1/\ta)^2$ and $\varphi$ is defined by $\cos(\varphi)=1/(\omega_0 \ta)$ and $\sin(\varphi) = \psi/\omega_0$.

Thus we observe experimentally that when we drive the system out of equilibrum, the angular fluctuations 
$\delta \theta$ are identical (with respect to the expressions above) to those at equilibrium. 
We verify the same properties for the sinusoidal time prescription of the torque, and use 
equilibrium expression for the correlation function in the next sections.

\subsection{Work distribution}
\label{sec:pdf_W}

In Figs.~\ref{fig:TFT}, \ref{fig:SSFTramp} and \ref{fig:sinuswork}, we see that the PDFs 
of the work are Gaussian for any integration time $\tau$ and 
whatever the forcing is. So these distributions are fully characterized by their 
mean value $\langle \Wt \rangle$ and their variance 
$\sigma_{\Wt}^2=\langle \delta \Wt^2 \rangle=\langle (\Wt - \langle \Wt \rangle)^2 \rangle$. 
The external torque $M$ is determistic, so the mean value of the work done on the system 
can be written as :
\begin{equation}
\langle \Wt \rangle = \frac{1}{ k_B T} \int_{t_i}^{t_i+\tau} \tilde{M}(t')\langle \dot{\theta}(t') \rangle \dd t'.
\label{eq:meangausswork}
\end{equation}
We have defined $\tilde{M}(t') = M(t') - a M(t_i)$. The value $a$ depends on the time-prescription
of the torque we apply to the oscillator. Choosing $a=1$ describes the linear ramp and $a=0$ 
corresponds to the sinusoidal forcing.

The variance of the PDFs is :
\begin{eqnarray}
\sigma_{\Wt}^2=\frac{1}{(k_B T)^2} \int_{t_i}^{t_i+\tau}\int_{t_i}^{t_i+\tau}  \tilde{M}(t_1).\tilde{M}(t_2)\langle \delta \dot{\theta}(t_2)\delta \dot{\theta}(t_1)\rangle \dd t_1\dd t_2.
\label{eq:vargausswork}
\end{eqnarray}
This expression involves the autocorrelation function of the angular speed $\delta \dot{\theta}$, 
($\langle \delta \dot{\theta}(t_2)\delta \dot{\theta}(t_1)\rangle$). Using the expression 
of the autocorrelation function of angular fluctuations ($R_{\delta \theta}$, 
Eq.~(\ref{eq:autocorrelation})), we can calculate exactly the expression of 
$\langle \delta \dot{\theta}(t_2)\delta \dot{\theta(t_1)}\rangle$:
\begin{equation}
\langle \delta\dot{\theta}(t_1)\delta\dot{\theta}(t_2) \rangle 
= -\,\frac{k_B T}{\Ieff\sin(\varphi)} \exp\left(-\frac{|t_2-t_1|}{\ta}\right) \sin(\psi|t_2-t_1|-\varphi).
\label{eq:autocorrdtheta}
\end{equation}
We have calculated the mean value and the variance of the PDFs in the three 
situations of interest: stationary and transient cases with a forcing linear in time,
and stationary case with a forcing sinusoidal in time. Details and results can be found
in the appendix. In all of the cases, we compare the theoretical PDFs and the symmetry 
functions with the experimental results. We have plotted in Fig.~\ref{fig:TFT}, 
\ref{fig:SSFTramp} and \ref{fig:sinuswork} our theoretical PDFs and the 
corresponding symmetry functions with no adjustable parameters. Within experimental 
error bars, our analytical and experimental results are in excellent agreement. 
$S(\wt)$ is linear in $\wt$ because the PDFs of the work are Gaussian. 
We now want to calculate analytically the corrections 
to the slope $\Sigma_w(\tau)$ for finite time $\tau$. 
For a Gaussian distribution, the symmetry function is :
\begin{equation}
S(\wt) = \frac{2\langle \Wt\rangle}{\sigma_W^2}\,\wt =\Sigma_w(\tau) \, \wt.
\label{eq:Gaussymfunc}
\end{equation}
The expression of the slope $\Sigma_w(\tau)$ uses only the mean value and the variance of the Gaussian distribution. 
We define $\Sigma(\tau)=(1-\epsilon(\tau))^{-1}$, where the correction $\epsilon(\tau)$ is a decreasing function 
of $\tau$. We obtain $\epsilon(\tau) = 0$ for the transient case, which is in agreement with a TFT. 
For the two steady states, there are corrections to the value $1$; we find:
\begin{enumerate}[i)]
\item{Linear forcing
\begin{equation}
\epsilon(\tau)=\frac{1}{\psi \tau}\left[\frac{A}{\omega_0 \tau}-\expdec \left(B+\frac{D}{\omega_0 \tau}\right)\right].
\label{eq:epsilon_linear}
\end{equation}}
\item{Sinusoidal forcing
\begin{equation}
\epsilon(\tau)=\frac{E}{\tau/\ta}+\frac{F}{\tau/\ta}\expdec.
\label{eq:epsilon_sinus}
\end{equation}}
\end{enumerate}
Exact values of the coefficients $A,B,D,E,F$ are given in the appendix. 
These two expressions are in perfect agreement with experimental results 
as can be seen in Fig.~\ref{fig:SSFTramp} and Fig.~\ref{fig:sinuswork}. 
These corrections depend on the kind of forcing and it is difficult to 
predict their form for an arbitrary forcing. Nevertheless, the two situations
we consider are useful as building blocks of such an arbitrary forcing, and 
they provide a very nice test of our method.

\section{Heat fluctuations : theoretical predictions\label{sec:heattheo}}

We now determine an analytical expression of the PDF of the dissipated heat. 
To do so, we make the same hypothesis as in the case of the work (see 
section~\ref{sec:exp_observations} above),
and we complete them by additional assumptions to simplify our derivations. 
We are interested by PDFs of the heat for integration time $\tau$ large compared 
to $\ta$, so that exponential corrections which are scaling like $\expdec$ can be neglected. 
In the case of sinusoidal forcing, this is correct after $3$ or $4$ periods of forcing 
($\tau/\ta=1.64n$). Within this assumption, $\theta(t_i+\tau)$ and $\theta(t_i)$ 
are independent, and so are $\frac{\dd \theta}{\dd t}(t_i+\tau)$ and 
$\frac{\dd \theta}{\dd t}(t_i)$. Additionally, as the equation of motion of the 
oscillator is second order in time, $\theta$ and $\frac{\dd \theta}{\dd t}$ 
are independent at any given time $t$. We use the technique proposed in~\cite{Cohen1}. 
To obtain the PDF $p(Q_\tau)$ of the heat, we define its Fourier transform, 
the characteristic function, as
\begin{equation}
\hat{P} _{\tau} (s) \equiv \int_{-\infty}^{\infty}  {\textrm d} q_\tau e^{isq_\tau} p(q_\tau)
\label{DefFourier}
\end{equation}
We then write $p(q_\tau)$ using Eq.~(\ref{eq:Qdef}) as:
\begin{equation}
p(q_\tau) = \int \int {\textrm d}\theta {\textrm d}\dot{\theta} \tilde{P}\left(\Delta U_\tau-Q_\tau, \theta(t_i+\tau) , \theta(t_i),\dot{\theta}(t_i+\tau), \dot{\theta}(t_i)\right).
\end{equation}
where $\tilde{P}$ is the joint distribution of the work $W_\tau$, $\theta$ and 
$\frac{{\textrm d}{\theta}}{{\textrm d}t}$ at the beginning and at the end of 
the time interval $\tau$. This distribution is expected to be Gaussian because 
$W_\tau$ is linear in $\dot{\theta}$ and additionally $\theta$, $\dot{\theta}$ 
and $W_\tau$ are  Gaussian. The details of the calculation are given 
in the appendix.

\subsection{Linear forcing}

The Fourier transform of the PDF of dissipated heat can be exactly calculated :
\begin{eqnarray}
\hat{P}_\tau(s) = \frac{1}{1+s^2} \exp
& \left\{ -d^2 i s\left(2 \frac{\tau}{\ta} + i s\left[2 \frac{\tau}{\ta} +2 \right] \right.\right. \nonumber \\
& + \left.\left. \frac{-16\cos(\varphi)^2+4+4 i s(4\cos(\varphi)^2+1)}{1+s^2}\right)\right\}.
\nonumber
\label{eq:FourierPDFheat_linear}
\end{eqnarray}
As far as we know, there is no analytic expression for inverse Fourier transform of this function, so for the PDF of dissipated heat. 
However we can do some comments. This expression is very similar to the one found in the case of a Brownian particle~\cite{Cohen1}. 
The factor $(1+s^2)^{-1}$ is the Fourier transform of an exponential PDF and this is directly connected to the exponential 
tails of the PDF. Moreover the PDF is not symmetric around its mean, because there is a non vanishing third moment. 
In this expression, only two terms depend on $\tau$. For large $\tau$, this expression reduces to:
\begin{eqnarray}
\hat{P}_\tau(s) = \frac{1}{1+s^2}
\exp\left\{-2id^2\frac{\tau}{\ta}s(1 + i s)\right\}.
\end{eqnarray}
This expression will turn out to be similar to the one obtain with a sinusoidal forcing, 
as we will comment in the next section.

Both expressions depend on the non-dimensional factor $d$ defined as:
\begin{equation}
d=\sqrt{\frac{1}{C k_B T}}\frac{M_o}{\omega_0 \tau_r} \,.
\end{equation}
All moments of the distribution of $Q_\tau$ are linear with $d^2$ and 
$\langle \dot{\theta} \rangle/\sqrt{\langle \delta \dot{\theta}^2 \rangle} = d$. 
So $d^2$ compares the mean value of the angular speed to the root mean square of 
the angular speed fluctuations. This coefficient $d^2$ increases when the system 
is driven further from equilibrium. We consider it as a measure of the distance 
to equilibrium. In our system $d$ is positive, but smaller than $1$, so we are 
out-of-equilibrium but not very far from it ($d = 0.059$).

\subsection{Sinusoidal forcing}
\label{sec:heat_sinus_theo}

Just like in the experiments, we choose the integration time $\tau$ to be a
multiple of the period of the forcing, so $\langle \Delta U_\tau \rangle = 0$ and therefore
$\langle W_\tau \rangle = - \langle Q_\tau \rangle$. Within this framework, we find that the PDF of $\Delta U_\tau$ is exponential:
\begin{equation}
P(\Delta U_\tau) = {1 \over 2} \exp (-|\Delta U_\tau|) \,.
\label{PDFdE}
\end{equation}
It is independant of $\tau$ because $\Delta U_\tau$ depends only on $\theta$ and $\frac{{\textrm d}{\theta}}{{\textrm d}t}$ at times $t_i$ and $t_i+\tau$ which are uncorrelated.
This expression is in perfect agreement with the experimental PDFs for all times (see Fig.~\ref{fig:sinusheat}b). 
Some algebra then yields for the characteristic function of $Q$:
\begin{equation}
\hat{P} _{\tau} (s)= \frac{1}{1+s^2}\exp \left( i\langle Q_\tau \rangle s-\frac{\sigma_W^2}{2} s^2 \right)
\label{eq:TFPDFQ}
\end{equation}
The characteristic function of heat fluctuations
is therefore the product of the characteristic function of an exponential distribution ($\frac{1}{1+s^2}$)
with the one of a Gaussian distribution ($\exp \left( i\langle Q_\tau \rangle s-\frac{\sigma_W^2}{2} s^2 \right)$).
Thus the PDF of heat fluctuations is nothing but the convolution of a Gaussian and an exponential PDF,
just as if $W_\tau$ and $\Delta U_\tau$ were independent. The inverse Fourier transform can be computed
exactly:
\begin{eqnarray}
P(Q_\tau) =  {1 \over 4}\exp \left( \frac{\sigma_W ^2}{2}\right) & \left[ e^{Q_\tau - \langle Q_\tau \rangle}
\textrm{Erfc} \left(\frac{Q_\tau-\langle Q_\tau \rangle+\sigma_W^2}{\sqrt{2\sigma_W^2}}\right) +
\nonumber \right.\\
&  \left. e^{-(Q_\tau - \langle Q_\tau \rangle)}\textrm{Erfc} \left(\frac{-Q_\tau+\langle Q_\tau \rangle+\sigma_W^2}{\sqrt{2\sigma_W^2}}\right)\right] \,,
\label{eq:PDFq}
\end{eqnarray}
where $\textrm{Erfc}(x)=1-\textrm{Erf}(x)$ stands for the complementary Erf function.
In Fig.~\ref{fig:sinusheat}c, we have plotted the analytical PDF from Eq.~(\ref{eq:PDFq}) together with the experimental ones,
using values of $\sigma_W^2$ and $\langle Q_\tau \rangle$ from the experiment and no adjustable parameters.
The agreement is perfect for all values of $n$, {\it i.e.} for any time $\tau$.
From Eq.~(\ref{eq:PDFq}), we isolate three different regions for $S(q_\tau)$:

(I) if $Q_\tau > \sigma_{W}^2 + |\langle Q_\tau \rangle|= 3 |\langle Q_\tau \rangle| + {\cal O}(1)$, then
$S(q_\tau) = 2+{\cal O}({1 \over \tau})$.
This domain of $S_\tau$ corresponds to fluctuations larger then three times the average value. 
The PDF has exponential tails, corresponding to an exponential distribution with a non-vanishing mean.

(II) if $Q_\tau < \sigma_{W}^2 - |\langle Q_\tau \rangle| = |\langle Q_\tau \rangle| + {\cal O}(1)$, then
$S(q_\tau)=\Sigma(n) q_\tau + {\cal O}({1 \over \tau})$ with 
$\Sigma_q(n) = \frac{2 |\langle Q_\tau \rangle|}{\sigma_W^2} = \Sigma_w(n) $. 
In this domain, values of the heat are small and heat fluctuations behave like work fluctuations. 
The slope $\Sigma(\tau)$ is the same as the one found for work fluctuations.
The exact correction to the asymptotic value 1 is plotted in Fig.~\ref{fig:sinusheat}e and again
it describes perfectly the experimental behavior.

(III) for $\sigma_{W}^2 - |\langle Q_\tau \rangle|  < Q_\tau < \sigma_{W}^2+ |\langle Q_\tau \rangle |$,
there is an intermediate region connecting domains (I) and (II) by a second order polynomial:
$S(q_\tau)=2 - {\Sigma (\tau) \over 4}(q_\tau-(1 + {2 \over \Sigma (\tau)}))^2+{\cal O}({1 \over \tau})$.

These three domains offer a perfect description of the three regions observed 
experimentally (Fig.~\ref{fig:sinusheat}d).

Now, we examine the limit of infinite $\tau$ in which SSFT is supposed to hold.
To do so, we distinguish two variables : the heat $Q_\tau$ or the normalized heat $q_\tau$.
Their asymptotic behavior are different because the average heat $\langle Q_\tau \rangle$
depends on $\tau$, more precisely it is linear in $\tau$.

We discuss first $Q_\tau$. The asymptotic shape of the PDF of $Q_\tau$ (Eq.~(\ref{eq:PDFq})) 
for large $\tau$ is a Gaussian whose variance is $\sigma_W^2$, the variance of the PDF 
of $W_\tau$. Thus, the PDF of $Q_\tau$ coincides with the PDF of $W_\tau$ for $\tau$ 
strictly infinite. As we have already shown, work fluctuations satisfy the conventional SSFT ; 
therefore heat fluctuations also satisfy the conventional SSFT (Eq.~(\ref{eq:SSFT})).
We have found three different regions separated by two limit values: the mean and three times the mean.
But in the limit of large times $\tau$, the PDF shrinks and only region (II) is relevant.
Region (II) corresponds to small fluctuations and it is bounded from above by 
$|\langle Q_\tau \rangle| + {\cal O}(1)$ with the average $\langle Q_\tau \rangle$
being linear in $\tau$. So all the behavior of the fluctuations of $Q_\tau$ for large $\tau$
lays in region (II) where the symetry function is linear and SSFT holds.

We turn now to the normalized heat $q_\tau$. As the average value of $Q_\tau$ is
linear in $\tau$, rescaling by $\langle Q_\tau \rangle$ is equivalent to a division by $\tau$ ;  
the mean of $q_\tau$ is then $1$. This normalization makes the two limit
values constant. The boundary between regions (II) and (III) is $1+{\cal O}(1/\tau)$ and 
the boundary between (III) and (I) is $3+{\cal O}(1/\tau)$.
The function $S(q_\tau)$ is not linear for large values of $q_\tau>1$ but 
it is linear only in region (II), for $q_\tau <1$, {\it i.e.} for small fluctuations.
So SSFT is satisfied only for small fluctuations but not for all values of $q_\tau$, 
and we obtain for $q_\tau$ a fluctuation relation which prescribes a symetry function that
is non-linear in $q_\tau$.

These two different pictures, in terms of $Q_\tau$ or $q_\tau$, results from taking 
two non-commutative limits differently. The first description using $Q_\tau$ implies 
that the limit $\tau$ infinite is taken before the limit of large $Q_\tau$. The second 
description does the opposite. However, the probability to have large fluctuations 
decreases with $\tau$ and experimentally, for large $\tau$, only the region (II) 
can be seen, and it is the region in which where SSFT holds.

As we have done in the case of the linear forcing, we introduce a non-dimensional factor $d$ such as :
\begin{eqnarray}
d&=& \sqrt{\frac{1}{C k_B T}} \frac{M_o\omega_d}{\omega_0 \rho(\omega_d))}\\ 
\rho(\omega_d)&=&\sqrt{\left(1-\left(\frac{\omega_d}{\omega_0}\right)^2\right)^2 + 4 \left( \frac{\omega_d}{\omega_0}\cos(\varphi)\right)^2}
\end{eqnarray}
The moments of the distribution of $\Qt$ are linear with $d^2$ and, like the linear torque, $d$ is equal to the amplitude of $\dot{\theta}$ divided by $\sqrt{\langle \delta \dot{\theta}^2 \rangle}$. We consider it also as a measure of the distance to equilibrium. In our system $d$ is positive, but smaller than $1$, so we are 
out-of-equilibrium but not very far it : here $d = 0.18$.

\section{Discussion and conclusion}
\label{sec:conc}

In this paper, we have studied the fluctuations of energy input and energy dissipation in a harmonic oscillator
driven out of equilibrium. This oscillator is very well described by a second order 
Langevin equation. We have performed experiments using a torsion pendulum driven out of equilibrium following a stationary protocol in which either the torque increases linearly in time, or oscillates at a given frequency. We have also studied 
transient evolutions from the equilibrium state. We have defined the work given to the system 
as the work of the torque applied during a time $\tau$. Accordingly we have defined the 
heat dissipated by the pendulum during this time $\tau$, by writing the first principle 
of thermodynamics between the two states separated by time $\tau$.

Fluctuations Relations are obtained experimentally for both the work and the heat, 
for the stationary and transient evolutions. 

We have experimentally observed that angle fluctuations of the brownian pendulum have the 
same statistical and dynamical properties at equilibrium and for any non-equilibrium
driving. From this observation, we have derived expressions for the probability density functions
of the work and the heat. In our system, fluctuations of the angle are Gaussian, and 
so are fluctuations of the work $\wt$. So the symmetry functions $S(\wt)$ of the work are 
linear, and we have calculated exactly the time-correction to the proportionality coefficient 
between $S(\wt)$ and $\wt$. These corrections match perfectly the experimental results, 
both in the case of a forcing linear in time, and sinusoidal in time. We have also computed
the analytic expression of the Fourier transform of the PDFs of the dissipated heat. 
For the sinusoidal forcing, we have obtained for the first time an analytic expression of 
the PDF of the heat. This expression is in excellent agreement with the experimental 
measurements. For a torque linear in time, the PDF of the heat 
has no simple expression but its Fourier transform gives insight on the behavior of 
the symmetry function of the heat. It is very similar to the one obtained in the case
of a first order Langevin dynamics~\cite{Cohen1}. We emphasize here that our analytical 
derivations are strongly connected to experimental observations on the properties of 
the noise ; and are therefore different from any previous theoretical approach. 

We have introduced a dimensionless variable $d$ which we think is a measure of the distance 
from equilibrium: the average dissipation rate is proportional to $d$, and it increases 
when the system is further from equilibrium. $d$ is also
proportional to the strength of the driving and in the Fluctuation Relations, it gives 
a proper unit to measure the amplitude of fluctuations. So $d$ plays the same role 
as the dissipation coefficient (the viscosity in our case) in the Fluctuation Dissipation 
Theorem at equilibrium. We have an expression of $d$ for the two different time-prescriptions we have used. These expressions can be generalized: 
\begin{equation}
d^2 = \frac{\langle \dot{\theta}^2 \rangle}{\langle \delta \dot{\theta}^2\rangle}
\end{equation}
The numerator corresponds to the solution of the Langevin equation when removing the thermal noise term ($\eta=0$). The denominator corresponds to the variance of thermal fluctuations of angular speed $\delta{\dot{\theta}}$.

We thank G. Gallavotti for useful discussions. This work has been partially supported 
by ANR-05-BLAN-0105-01.

\appendix
\addcontentsline{toc}{section}{Appendix}

\section*{Work fluctuations}
\addcontentsline{toc}{section}{A	Work fluctuations}
In this section, we will calculate the mean and the variance of the work given to the system in the cases :
\begin{enumerate}
\item{Transient state, linear forcing}
\item{Steady state, linear forcing}
\item{Steady state, sinusoidal forcing}
\end{enumerate}

\subsection*{TFT, forcing linear in time}
\addcontentsline{toc}{subsection}{A.1	TFT, forcing linear in time}
\label{TFTramp}
The torque is : $M(t) = M_o \, t/\tau_r$. The mean value of the angular 
displacement is the solution of Eq.~\ref{eq:Langevin_oscillator} :
\begin{equation}
\langle \theta \rangle = \frac{M_o}{\psi C \tau_r} \left( e^{-t/\ta}\sin(\psi t + 2 \varphi)  + \psi t - \sin(2 \varphi)\right).
\label{meanthetatransient}
\end{equation}
For the work done on the system, the PDFs are Gaussian for all integration time $\tau$. 
The mean of the PDF of $W_\tau$ for a given $\tau$ is:
\begin{eqnarray}
\langle \Wt \rangle = \frac{M_o^2}{k_B T\psi C \tau_r^2}
& \left[\frac{1}{2}\psi \tau^2 +\tau \expdec\sin(\psi \tau+2\varphi)  		\nonumber \right.\\
& \left. +\frac{1}{\omega_0}\left(\expdec\sin(\psi \tau+3\varphi)-\sin(3\varphi)\right)\right]
\label{eq:meanworkramp}
\end{eqnarray}
and its variance is:
\begin{eqnarray}
\sigma_{\Wt}^2 = \frac{2 M_o^2}{k_B T \psi C \tau_r^2}
& \left[\frac{1}{2}\psi \tau^2 +\tau \expdec\sin(\psi \tau+2\varphi)  \nonumber \right.\\
& \left. + \frac{1}{\omega_0}\left(\expdec\sin(\psi \tau+3\varphi)-\sin(3\varphi)\right)\right],
\label{eq:varianceworkramp}
\end{eqnarray}

\begin{eqnarray}
\sigma_{\Wt}^2 = 2 \langle \Wt \rangle\,.
\end{eqnarray}
\subsection*{SSFT, forcing linear in time}
\addcontentsline{toc}{subsection}{A.2	SSFT, forcing linear in time}
\label{SSFTramp}
The torque is : $M(t) = M_o \, t/\tau_r$. The mean value of the angular displacement is the solution of Eq.~\ref{eq:Langevin_oscillator} after some $\ta$. Thus the exponential term is vanished:
\begin{equation}
\langle \theta \rangle = \frac{M_o}{\psi C \tau_r} \left(\psi t - \sin(2 \varphi)\right).
\label{meanthetastat}
\end{equation}
For the work done on the system, the PDFs are Gaussian for all integration time $\tau$. The mean of the PDF is:
\begin{equation}
\langle \Wt \rangle= \frac{M_o^2}{2 k_B T C\,\tau_r^2} \tau^2
\label{eq:meanworkrampstat}
\end{equation}
and the variance is:
\begin{eqnarray}
\sigma_{\Wt}^2 = \frac{2 M_o^2}{k_B T \psi C \tau_r^2}&\left[\frac{1}{2}\psi \tau^2 +\tau \expdec\sin(\psi \tau+2\varphi)  \nonumber \right.\\
&\left. + \frac{1}{\omega_0}\left(\expdec\sin(\psi \tau+3\varphi)-\sin(3\varphi)\right)\right].
\label{eq:varianceworkrampstat}
\end{eqnarray}
From this, we deduce:
\begin{eqnarray}
\epsilon(\tau)&=&\frac{1}{\psi \tau}\left\{A-\expdec\left(B+\frac{D}{\omega_0\tau}\right)\right\}\,,
\end{eqnarray}
where
\begin{eqnarray}
A&=& 2\frac{\sin(3\varphi)}{\omega_0\tau}\,, \nonumber\\
B&=& 2\sin(\psi\tau+2\varphi)\nonumber\,,\\
D&=&2\sin(\psi\tau+3\varphi)\nonumber\,.
\end{eqnarray}

\subsection*{SSFT, forcing sinusoidal in time}
\addcontentsline{toc}{subsection}{A.3	SSFT, forcing sinusoidal in time}
\label{SSFTsinus}
The torque is $M(t)=M_o\,\sin(\omega_d t)$. The mean value of the angular displacement is:
\begin{eqnarray}
\langle \theta \rangle=\theta_0 \sin(\omega_d t+\beta) \quad \textrm{where}\quad \theta_0=\frac{M_0}{C \,\rho(\omega_d)}
\end{eqnarray}
where
\begin{eqnarray}
\cos(\beta)&=&\frac{1-\left(\frac{\omega_d}{\omega_0}\right)^2}{\rho(\omega_d)}
\quad {\mathrm{and}}\quad
\sin(\beta)=\frac{-2\left(\frac{\omega_d}{\omega_0}\right)\cos(\varphi)}{\rho(\omega_d)}
\nonumber \,,\\ 
\rho(\omega_d)&=&\sqrt{\left(1-\left(\frac{\omega_d}{\omega_0}\right)^2\right)^2 + 4 \left( \frac{\omega_d}{\omega_0}\cos(\varphi)\right)^2}\,.
\end{eqnarray}
For the work done on the system, the PDFs are Gaussian for all integration time $\tau$. The mean of the PDF is:
\begin{equation}
\langle \Wn\rangle=\frac{M_o^2}{k_B T C}\left(\frac{\omega/\omega_0}{\rho(\omega)}\right)^2(\tau/\ta)
\label{eq:SSFT_sinus_mean_PDF_work}
\end{equation}
and the variance is:
\begin{eqnarray}
\sigma_n^{2}=2 \langle \Wn \rangle +E+F\expdec 
\end{eqnarray}
where
\begin{eqnarray}
E=-\frac{\langle W_n \rangle (1+(\omega/\omega_d)^2)\cos(2\beta)}{(\omega/\omega_0)^2(\tau/\ta)}
\end{eqnarray}
\begin{eqnarray}
F=-\frac{\langle W_n \rangle}{(\omega/\omega_0)^2.(\tau/\ta))} & \left[\sin(\psi \tau+\varphi)\cos(2\beta)+(\omega/\omega_0)^2 \sin(\psi\tau-\varphi)\cos(2\beta)\right.\nonumber\\ & \left.+(\omega/\omega_0)\sin(\psi \tau)\sin(2\beta)\right]\,.\label{eq:SSFT_sinus_Var_PDF_work}
\end{eqnarray}

\section*{Heat fluctuations}
\addcontentsline{toc}{section}{B	Heat fluctuations}
\label{Appendixheat}
In this section, we will calculate the Fourier transform of the PDF of the dissipated heat in two cases :
\begin{enumerate}
\item{Linear forcing}
\item{Sinusoidal forcing}
\end{enumerate}

\subsection*{Linear forcing}
\addcontentsline{toc}{subsection}{B.1	Linear forcing}
We introduce non-dimensional parameters in order to simplify calculations :
\begin{eqnarray}
\tilde{x}(t)&=&\sqrt{\frac{C}{k_B T}}\left(\theta(t)-\frac{M(t)}{C}\right),\nonumber\\
\dot{x}&=&\sqrt{\frac{\Ieff}{k_B T}}\dot\theta(t). \label{eq:nondim_var_ramp_heat}
\end{eqnarray}
The mean value and the variance of $\tilde{x}$ and $\dot{x}$ can be simply expressed :
\begin{eqnarray}
d&=&\sqrt{\frac{M_o^2}{C k_B T}}\frac{1}{\omega_0 \tau_r},\nonumber\\
\langle \tilde{x} \rangle &=& -2.d.\cos (\varphi)\,,\langle \dot{x} \rangle = d\,, \langle \delta \tilde{x}^2 \rangle =1\,,\langle \delta \dot{x}^2 \rangle =1.
\end{eqnarray}
where $d$ is a non-dimensional value. Integrating by part, the work $W_\tau$ can be rewritten:
\begin{eqnarray}
W_\tau&=& d.\omega_0\left[(t_i+\tau) x(t_i+\tau)-t_i x(t_i)\right]-\frac{(d\omega_0)^2}{2}\left[(t_i+\tau)^2-t_i^2\right]+W^{*},\nonumber\\
W^{*} &=& -(d\omega_0)\int_{t_i}^{t_i+\tau}\tilde{x}(t')\dd t'.
\end{eqnarray}
With these definitions, we obtain : $Q_\tau = \frac{1}{2}\Delta \tilde{x}_\tau+\frac{1}{2}\Delta \dot{x}_\tau - W^{*}$ and $\langle Q_\tau \rangle = - \langle W^{*} \rangle$.
Like the distribution of $W_\tau$, the distribution of $W^{*}$ is gaussian for all values of $\tau$ and we find:
\begin{eqnarray}
\langle W^{*} \rangle = 2 d^2 \tau/\ta,\nonumber\\
\sigma_{W^{*}}^2=2 d^2 \left(2 \tau/\ta+1-4 \cos(\varphi)^2\right).
\end{eqnarray}
For notational convenience, we introduce a five dimensional vector : $Y = (W^{*}, \tilde{x}(t_i+\tau), \tilde{x}(t_i), \dot{x}(t_i+\tau), \dot{x}(t_i))$. $\tilde{P}$ is Gaussian and is so fully characterized by the covariance matrix ${\cal C}$ defined as:
\begin{equation}
{\cal C}_{ij}=\langle (Y_i-\langle Y_i\rangle)(Y_j-\langle Y_j \rangle)^{\dag} \rangle
\end{equation}
where $Z^{\dag}$ denotes the complex conjugate of $Z$. So the distribution $\tilde{P}$ is written :
\begin{equation}
\tilde{P}\left(Y\right)=\sqrt{\frac{1}{(2 \pi)^5 \mathrm{det}{\cal C}}}\exp\left(-\frac{1}{2} (Y-\langle Y \rangle)^T C^{-1} (Y-\langle Y \rangle)\right)\end{equation}
where $Z^T$ denotes the transpose of $Z$. We suppose that the integration time is larger than the relaxation time. Within this assumption, $\theta(t_i+\tau)$ and $\theta(t_i)$ are independent, and so are $\dot{\theta}(t_i+\tau)$ and $\dot{\theta}(t_i)$. As the equation of motion of the oscillator is second order in time, $\theta$ and $\dot{\theta}$ are independent at any given times $t$. With these hypotheses, we get :
\begin{eqnarray}
\langle \delta\tilde{x}(t_i)\delta\tilde{x}(t_i+\tau)\rangle &=&\langle \delta \tilde{x}(t_i) \delta\dot{x}(t_i) \rangle = \langle \delta \tilde{x}(t_i)\delta\dot{x}(t_i+\tau) \rangle \nonumber\\
&=& \langle \delta\dot{x}(t_i)\delta \tilde{x}(t_i+\tau)\rangle = \langle \delta\dot{x}(t_i)\delta\dot{x}(t_i+\tau) \rangle \nonumber\\ &=& \langle \delta\dot{x}(t_i+\tau)\delta\tilde{x}(t_i+\tau)\rangle = 0.
\end{eqnarray}
The other coefficients of the covariance matrix are :
\begin{eqnarray}
\langle \delta W^{*} \delta\tilde{x}(t_i) \rangle &=& \langle \delta W^{*} \delta\tilde{x}(t_i+\tau) \rangle = - 2 d \cos(\varphi),\nonumber\\
\langle \delta W^{*} \delta\dot{x}(t_i) \rangle &=& \langle \delta W^{*} \delta\dot{x}(t_i+\tau) \rangle = -  d.
\end{eqnarray}

The Fourier transform of the PDF of the heat will be calculated here. We define two quantities : 
\begin{eqnarray}
\vec{e}=\left(\begin{array}{c}
1\\ 0\\ 0\\ 0\\ 0
\end{array}\right)
\qquad N=
\left(\begin{array}{ccccc}
0&0&0&0&0\\
0&1&0&0&0\\
0&0&-1&0&0\\
0&0&0&1&0\\
0&0&0&0&-1
\end{array}\right).
\end{eqnarray}
One can write $Q_\tau$ as : $Q_\tau = \frac{1}{2}Y^T N Y-e^T Y$. The Fourier transform can be so written :
\begin{eqnarray}
\hat{P}_\tau(s)&=&\int \frac{\dd Y}{\sqrt{(2\pi)^5\mathrm{det}{\cal C}}} \exp\left(M\right),\nonumber\\
M&=&-\frac{1}{2}(\delta Y)^T {\cal C}^{-1} (\delta Y) + i \, s \left(\frac{1}{2}Y^T N Y-e^T Y\right).
\end{eqnarray}
We use a new variable defined as :
\begin{equation}
Y'=Y - (1-i\, s {\cal C}.N)^{-1}(\langle Y \rangle-i s.{\cal C}.e).
\end{equation}
With this definition, the argument in the exponential $M$ can be rewritten :
\begin{eqnarray}
M&=&-\frac{1}{2} Y' ({\cal C}^{-1}-i s N) Y'+\gamma,  \nonumber \\
\gamma&=&\frac{i \, s}{2}\left[(N\langle Y\rangle-e)^T(1-i \, s {\cal C}N)^{-1}(\langle Y \rangle - i \, s {\cal C}e)-\langle Y\rangle^T e\right].
\end{eqnarray}
Changing the integration variable to $Y'$ yields :
\begin{eqnarray}
\hat{P}_\tau(s) &=& \int \frac{\dd Y'}{\sqrt{(2\pi)^5\mathrm{det}{\cal C}}}\exp\left(-\frac{1}{2} Y'^{T}({\cal C}^{-1}-i \, s N)Y'\right) . \exp (\gamma)\nonumber\\
&=&\frac{\exp (\gamma)}{\sqrt{\mathrm{det}(1-i\, s {\cal C}.N)}}.
\label{eq:gen_heat_lin_exp}
\end{eqnarray}
To make Eq.~\ref{eq:gen_heat_lin_exp} into an explicit expression for $\hat{P}_\tau$, the inverse of matrix $(1-i s {\cal C}.N)$ is required in the expression for $\gamma$ and its determinant. These are obtained as follows. We find:
\begin{eqnarray}
1-i \, s {\cal C}.N = \left(\begin{array}{ccccc}
1&i\,s(2d\cos(\varphi))&i\,s(-2d\cos(\varphi))&i\,s d& -i\,sd\\
0&1-i\,s&0&0&0\\
0&0&1+i\,s&0&0\\
0&0&0&1-i\,s&0\\
0&0&0&0&1+i\,s
\end{array}\right).\nonumber
\end{eqnarray}
The determinant of the matrix is $(1+s^2)^2$. For the inverse of this matrix, we get : 
\begin{eqnarray}
(1-i \,s {\cal C}.N)^{-1} = \left(\begin{array}{ccccc}
1&-\frac{i\,s}{1-i\,s}(2d\cos(\varphi))&\frac{i\,s}{1+i\,s}(2d\cos(\varphi))&-\frac{i\,s}{1-i\,s} d& \frac{i\,s}{1+i\,s}d\\
0&\frac{1}{1-i\,s}&0&0&0\\
0&0&\frac{1}{1+i\,s}&0&0\\
0&0&0&\frac{1}{1-i\,s}&0\\
0&0&0&0&\frac{1}{1+i\,s}
\end{array}\right).\nonumber
\end{eqnarray}
We now have the material needed to calculate $\gamma$. We find : 
\begin{eqnarray}
\gamma=-i\,s \langle W^* \rangle -\frac{s^2}{2}&\left[\sigma_W^2+2 d^2(1+4\cos(\varphi)^2) +\frac{4 i\,s d^2}{1+s^2}(4\cos(\varphi)^2-1) \right.\nonumber\\ &\left.+\frac{4s^2d^2}{1+s^2}(4\cos(\varphi)^2+1)\right].
\end{eqnarray}
The analytic expression of the Fourier transform of the PDF of the heat dissipated during a linear forcing is :
\begin{eqnarray}
\hat{P}_\tau(s) = \frac{1}{1+s^2} \exp
& \left\{ -d^2 i s\left(2 \frac{\tau}{\ta} + i\, s\left[2 \frac{\tau}{\ta} +2 \right] \right.\right. \nonumber \\
& + \left.\left. \frac{-16\cos(\varphi)^2+4+4 i \,s(4\cos(\varphi)^2+1)}{1+s^2}\right)\right\}.
\end{eqnarray}

\subsection*{Sinusoidal forcing}
\addcontentsline{toc}{subsection}{B.2	Sinusoidal forcing}
\label{ap:sinus_heat}
We will determine in a first time the Gaussian joint distribution $\tilde{P}$ of $W_\tau$, $\theta(t_i)$, $\theta(t_i+\tau)$, $\dot{\theta}(t_i)$ and $\dot{\theta}(t_i+\tau)$. For notational convenience, we introduce a five dimensional vector : $\vec{X} = (W_\tau, \theta(t_i+\tau), \theta(t_i), \dot{\theta}(t_i+\tau), \dot{\theta}(t_i))$. The PDF $\tilde{P}$ is fully characterized by the covariance matrix ${\cal C}$.
\begin{equation}
{\cal C}_{ij}=\langle (X_i-\langle X_i\rangle)(X_j-\langle X_j \rangle)^\dag \rangle
\end{equation}
where $Z^\dag$ denotes the complex conjugate of $Z$. We suppose that the integration time is larger than the relaxation time. Within this assumption, $\theta(t_i+\tau)$ and $\theta(t_i)$ are independent, and so are $\dot{\theta}(t_i+\tau)$ and $\dot{\theta}(t_i)$. As the equation of motion of the oscillator is second order in time, $\theta$ and $\dot{\theta}$ are independent at any given times $t$. With these hypotheses, we get :
\begin{eqnarray}
\langle \delta\theta(t_i)\delta\theta(t_i+\tau)\rangle &=&\langle \delta\theta(t_i) \delta\dot{\theta}(t_i) \rangle = \langle \delta \theta(t_i)\delta\dot{\theta}(t_i+\tau) \rangle \nonumber\\
&=& \langle \delta\dot{\theta}(t_i)\delta \theta(t_i+\tau)\rangle = \langle \delta\dot{\theta}(t_i)\delta\dot{\theta}(t_i+\tau) \rangle \nonumber\\ &=& \langle \delta\dot{\theta}(t_i+\tau)\delta\theta(t_i+\tau)\rangle = \langle \Wt \theta(t_i) \rangle \nonumber\\ &=& \langle \Wt \theta(t_i+\tau) \rangle = \langle \Wt \dot{\theta}(t_i) \rangle  \nonumber\\ &=& \langle \Wt \dot{\theta}(t_i+\tau) \rangle = 0.
\end{eqnarray}
The covariance matrix is so a diagonal matrix :
\begin{eqnarray}
{\cal C} = \left( \begin{array}{ccccc} \sigma_W^2 & 0 &0 &0 &0 \\ 0& k_B T/C &0&0&0\\
0&0&k_B T/C&0&0\\ 0&0&0&k_B T/\Ieff &0\\ 0&0&0&0&k_B T/\Ieff \end{array}\right).
\end{eqnarray}
$\DUt$ is a function of the positions and velocities at the beginning ($t_i$) and at the end ($t_i+\tau$). Thus, $\DUt$ and $\Wt$ can be considered as independent. The PDF of $\Qt$ is so the convolution between the PDF of $\Wt$ which is Gaussian and the PDF of $\DUt$ :
\begin{equation}
P(\Qt) = \int_{-\infty}^{+\infty} P_{\Wt}(z)P_{\DUt}(\Qt+z)\dd z.
\end{equation}

We first calculate exactly the PDF of the variation of internal energy. We have shown that $\theta(t_i)$, $\theta(t_i+\tau)$, $\dot{\theta}(t_i)$ and $\dot{\theta}(t_i+\tau)$ are independent. The Fourier transform of the PDF is :
\begin{equation}
\hat{P}_{\DUt} (s) = \hat{P}_{E_p(t_i+\tau)}(s).\hat{P}_{E_c(t_i+\tau)}(s).\hat{P}_{E_p(t_i)}(-s).\hat{P}_{E_c(t_i)}(-s) 
\end{equation}
where $E_p = \frac{1}{2 k_BT} C \theta^2$ and $E_c = \frac{1}{2k_BT} \Ieff \dot{\theta}^2$. The distribution of $\theta$ is gaussian with variance $k_B T/C$. The distribution of $E_p$ and the distribution of $E_c$ are the same :
\begin{equation}
P_{E_p}(x) = P_{E_c}(x) = \frac{1}{\sqrt{\pi x}}\exp(-x).
\end{equation}
The Fourier transform of this distribution is : $\hat{P}(s) = (1-i s)^{-1/2}$. This distribution is the same for $E_p$ and $E_c$ at $t_i$ and $t_i+\tau$. Thus the Fourier transform of the variation of internal energy is 
\begin{equation}
\hat{P}_{\DUt}(s) = (1+s^2)^{-1}
\end{equation}
and the probability is : 
\begin{equation}
P(\DUt) = \frac{1}{2}\exp(-|\DUt |).
\end{equation}

As $\DUt$ and $\Wt$ are independent, the Fourier transform of the dissipated heat can be calculated :
\begin{equation}
\hat{P}_{\Qt}(s) = \frac{\exp\left(i\langle\Qt\rangle-\frac{\sigma_W^2}{2}s^2\right)}{1+s^2}.
\end{equation}
This expression can be inversed because it is simply the convolution between a Gaussian distribution 
and an exponential distribution. So we find Eq.~(\ref{eq:PDFq}).

\end{document}